# Design of quantum correlation based automatic focus detection system during dental laser operation

Xuan Binh Cao[1*], Duc An Pham[1]

[1] School of Mechanical Engineering, Hanoi University of Science and Technology, Hanoi 100000, Vietnam

**Abstract:** Dental laser has been developed recently and gradually replaced the conventional dental treatment methods, especially in dental caries removal. The utility of laser in dentistry contributes to reduce the pain during surgery and decrease the risk of sequelae after surgery. Although bringing about several beneficial properties such as excellent directionality, high speed, ultra-tiny active area and tunable energy concentration, laser is still quite limited in dental surgery because of the difficulty in real-time high-precision detection of laser focus during the operation. In this study, a cutting-edge optical system for real-time focus detection during dental laser operation employing the quantum entanglement of photon pairs created via spontaneous parametric down-conversion (SPDC) is presented. Defocusing distance of the specimen which is used as the sample tooth can be determined precisely through spatial correlation of photons arriving at the detectors. The proposed technique can be applied to design automatic focus detection system for use in dental laser operation.

## 1    Introduction:

Laser technology is widely applied in several fields of medicine such as medical imaging [1], angioplasty, dermatology, lithotripsy [2], dentistry [3-8], owning to the excellent orientation, rapid pulses, small active area, low noise, and changeable energy concentration. Particularly, dental laser has been employed for caries removal in which the tiny focal spot of high intensity laser is located on the target surface of the enamel to remove the dental decay caused by bacteria [9-12]. The method is expected to be a potential candidate alternating the traditional rotating mechanical drill for treatment of dental decay with purpose of reduction of pain, side effects, and vibration. However, in order to complete this mission, it is the prerequisite to have a high-precision real-time focus detection which are capable for precisely positioning the laser focal spot on the target caries. Real-time focal position detection plays important roles in a wide range of laser applications in medicine. A small misalignment between the target sample surface and the focus can have several consequences. First, if the laser intensity is not properly concentrated on the sample surface, the healthy tissues will be destroyed, and post-surgery sequelae will be expected to occur. Second, numerous side effects such as scarring, bleeding, and infection can occur on the tissue surface, possibly damaging the surface permanently. Therefore, focal position identification during laser operation is extremely important and necessary for both scientists and doctors in the medical field, particularly those who work with high-intensity lasers in dentistry.

Numerous methods have been proposed to optimize focus determination [13–21]. However, it is difficult to perform laser operation and detection simultaneously with most of these techniques. In addition, the laser source used for detection is usually not sufficiently powerful to enable cutting operation on tough materials such as hard tissues. Moreover, some methods require the use of complicated optical elements and auxiliary laser sources for focus detection. Quantum correlation has recently demonstrated to be a high-precision and sensitive concept of optical metrology [22–25]. This method enables the measurement of distances even smaller than the scale of the beam waist, making it highly effective for identifying defocusing distances from target sample surfaces of arbitrary roughness during laser fabrication. Employing quantum spatial correlation, the average of the positions of the two correlated photons coming in the detectors is characterized, rather than the positions of the independent single photons. This concept appears to be incredibly more precise and reliable than classical techniques of uncorrelated photons measurement. Quantum entanglement has a broad range of applications in science and technology [26–36]. Although it possesses widely known accuracy down to the Heisenberg scale [37,38], its practical uses in high-precision laser surgery as well as the laser medicine seem to be peculiar. Therefore, an automatic optical quantum focusing equipment employing biphoton spatial correlation in dental laser operation for caries removal is expected to provide benefits of precision and aesthetics.

This letter presents a novel method of positioning a target enamel sample (specimen), which provide the perfect reflectivity, at the focus during dental laser operation for caries removal by utilizing the quantum correlation of the photon positions on the detectors as well as examining the beam spot size on the image sensor. A simple optical system is employed in this method to detect the focus with a high accuracy and low noise and can be used to measure small discrepancies between the target sample surface and the focus. The setup includes two subsystems. One subsystem performs focus detection using the geometrical shape of the beam spot while the specimen is shifted along the optical axis. The other subsystem measures the defocusing distance on the basis of the quantum entanglement of the biphoton positions by counting the numbers of correlated photons reaching the left and right sides of the detectors. It also generates the probability distribution of the positions of the entangled photons while the stream of correlated photons is deflected as the specimen is moved owing to the fact that the reflecting mirror and specimen are located on the same micro-positioning stage. The defocusing distance of the specimen is exactly equal to the shift in the photon stream. The system does not require an auxiliary laser source for focus detection during operation.

## 2 Methodology:

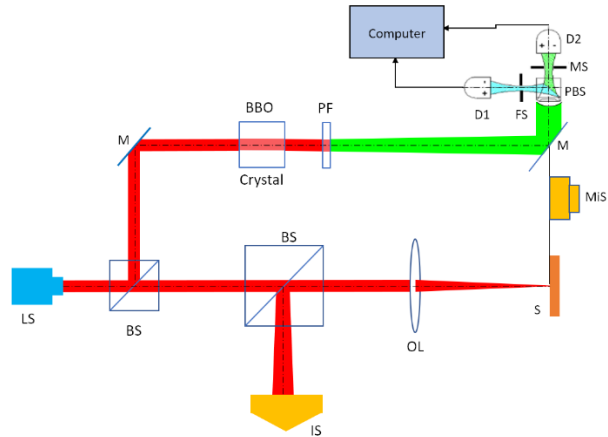

Fig. 1. The symbolistic description of optical system used for focus detection based on biphoton spatial correlation. The first fractional beam conducts optical displacement measurement. This photon stream is sent through a barium borate (BBO) crystal. This crystal converts a photon into two entangled photons with orthogonal polarizations via spontaneous parametric down-conversion (SPDC). The two entangled photons get reflected by the mirror (M) and separated by a polarized beam splitter (PBS) before hitting detectors D1 and D2 at the transverse positions $x_1$ and $x_2$, respectively. The uncorrelated elements are removed by pump filter (PF), and the fractional correlated phton streams are focused at the fixed slit (FS) and movable slit (MS), where the correlation distribution width are manipulated [26,27]. In this method, the dependences of the correlated position distribution on the variances of $x_1 + x_2$ and $x_1 - x_2$ are examined to determine the defocusing distance of the specimen which is used to replace the real enamel surface (S). S and M, whose positions are controlled by a microstage (MiS), are connected. The second fractional beam is directed toward the specimen through the objective lens (OL) and is reflected toward the image sensor (IS) by another beam splitter (BS). The configuration of the beam spot is investigated to determine the focal position of the specimen, which is the reference point for the defocusing distance measurements.

The optical system can be described in detail as follows. First, the laser beam from the source is split into two fractional beams by a beam splitter. One fractional beam is directed through the first subsystem, which is composed of a specimen, an objective lens, a beam splitter, and an image sensor. The specimen used for replacement of enamel sample in this technique is a crystalline calcium phosphate sample with the perfect reflectivity that is movable along the optical axis, enabling modeling of a target sample surface with an arbitrary roughness. The fractional beam passes through the objective lens, hits the specimen surface, is reflected, passes through the objective lens again, is reflected by the beam splitter, and finally arrives at the image sensor. Obviously, the beam spot size on the image sensor depends on the position of the specimen with respect to the objective lens. Using this principle, the focal position can be easily detected on the basis of the beam spot size on the image sensor, which is designed to read the beam spot radius. To perform this detection, the first fractional laser beam is prevented from reaching the objective lens by a blocking plate. The laser beam is then only reflected by the beam splitter and hits the image sensor, and the reflected beam spot radius is recorded. Next, the blocking plate is removed and the beam spot radius is recorded while shifting the specimen along the optical axis until the beam spot radius achieves the value recorded during the first step. The position at which the same beam spot radius is obtained is the focal position of the specimen. By performing this calibration, the reference point for measuring the defocusing distance during dental laser operation can be identified. The detailed calibration and focus detection were presented in our previous report [39,40].

However, when laser fabrication is performed on a real target sample, the sample surface is not always located at the focus, decreasing the fabrication quality. The focus detection system should automatically identify the defocusing state and measure the defocusing distance of the sample surface precisely. Therefore, the second fractional beam from the laser source or pump beam is directed through a nonlinear medium (e.g., a barium borate (BBO) crystal) experiencing spontaneous parametric down-conversion (SPDC). Specifically, the incident single photon from the second fractional beam is converted into a pair of outgoing photons via SPDC in the BBO crystal. These two photons are correlated owing to conservation of energy and momentum and express their correlation through the probability distribution of their positions on the detectors. They are split by a polarized beam splitter and directed toward two detectors that are photon counters. The correlated position distribution varies according to the distance of the reflecting mirror, which is set on the same microstage as the specimen, from the focal position. Consequently, the defocusing displacement of specimen is given by [25–27].

$$\Delta = \frac{x_1 + x_2}{2} = \sqrt{\frac{\pi(\sigma^2 + \epsilon^2)}{8}} \frac{N_+ - N_-}{N_+ + N_-}. \tag{1}$$

where $\sigma$ is the waist of the pump beam, which is calculated using $\sigma = \sqrt{9L\lambda_o/10\pi}$, where $L$ is the thickness of the crystal, and $\lambda_o$ is the wavelength of the pump beam; $x_1$ and $x_2$ are the transverse positions of the photons in a given pair on detectors 1 and 2, respectively; and $\epsilon$ is the spatial correlation parameter or pump radius at the BBO crystal, which is calculated using $\epsilon = w_o\sqrt{1 + \left(\frac{z}{z_R}\right)^2}$, where $z_R = \frac{\pi w_o^2}{\lambda_o}$ is the Rayleigh length, $w_o$ is laser beam waist at the laser source, and $z$ is the distance along the beam path between the laser source and the exit face of the nonlinear crystal. We define $N_-$ and $N_+$ as the numbers of photons reaching the left and right sides of the detectors, respectively, where $N = N_- + N_+$ is the total number of photons. Var($\Delta$) indicates the accuracy of the $\Delta$ measurement which is calculated by [25-27]:

$$\text{Var}(\Delta) = \frac{(\epsilon^2 + \sigma^2)(\pi + 2\arcsin\xi)}{16N}. \tag{2}$$

where $\xi = \frac{\epsilon^2 - \sigma^2}{\epsilon^2 + \sigma^2}$ is the entanglement coefficient between two correlated photons. Obviously, when the total number of photons arriving at the detectors increases, the measurement uncertainty decreases. In other words, when we conduct the measurement with higher laser power, the accuracy of focus detection increases. This demonstrates the suitable integration of the detection setup with high power pulsed laser system used in laser surgery. In the setup, the detectors D1 and D2 are the combination of position-sensitive detectors (PSD) and avalanche photo-diodes (APD). The number of detector pixels is 50 with detection efficiency of 85% and thickness of BBO is 2 mm. The errors in detection efficiency, beam waist, deviation of the beam due to misalignment, and repetition rate are ±0.3 %, ±0.4 %, ±0.1 %, ±0.1 % respectively.

## 3. Discussion

As an engineer, one may also be interested in calculating the beam spot radius, which depends on the positions of the optical elements along the optical axis. The analytical expression for this radius can facilitate estimation of the number indicated by the image sensor and detection of the focal position along with calibration. First, to determine the dependencies of the beam spot size on the image sensor, we examined the optical pathway of the laser beam from the laser source to the image sensor and employed a ray transfer matrix analysis. Here, $\begin{pmatrix} 1 & \rho \\ 0 & 1 \end{pmatrix}$, $\begin{pmatrix} 1 & 2(f + \Delta) \\ 0 & 1 \end{pmatrix}$, and $\begin{pmatrix} 1 & a \\ 0 & 1 \end{pmatrix}$ are the transfer matrices for the propagation of the laser beam in free space, and $\begin{pmatrix} 1 & 0 \\ -f^{-1} & 1 \end{pmatrix}$ is the transfer matrix for the propagation of the laser beam through a thin lens. Therefore,

$$T = \begin{pmatrix} 1 & \rho \\ 0 & 1 \end{pmatrix} \cdot \begin{pmatrix} 1 & 0 \\ -f^{-1} & 1 \end{pmatrix} \cdot \begin{pmatrix} 1 & 2(f + \Delta) \\ 0 & 1 \end{pmatrix} \cdot \begin{pmatrix} 1 & 0 \\ -f^{-1} & 1 \end{pmatrix} \cdot \begin{pmatrix} 1 & a \\ 0 & 1 \end{pmatrix}$$

$$= \begin{pmatrix} \frac{2\Delta}{f^2}(\rho - f) - 1 & \frac{2\Delta}{f^2}(a - f)(\rho - f) + 2f - a - \rho \\ \frac{2\Delta}{f^2} & \frac{2\Delta}{f^2}(a - f) - 1 \end{pmatrix} = \begin{pmatrix} A & B \\ C & D \end{pmatrix}. \tag{3}$$

For the input beam, the complex radius of curvature $q_1$ is

$$q_1 = iz_R \tag{4}$$

and

$$q = \frac{1}{\frac{1}{R} - \frac{i\lambda_o}{\pi w^2}} = \frac{A \cdot q_1 + B}{C \cdot q_1 + D} = \frac{(2\Delta f^{-2}(\rho - f) - 1)(i z_R) + 2\Delta f^{-2}(f - a)(f - \rho) + 2f - a - \rho}{2\Delta f^{-2}(i z_R) + 2\Delta f^{-2}(a - f) - 1} \tag{5}$$

$$\to \frac{1}{R} - \frac{i\lambda_o}{\pi w^2} = \frac{\{4\Delta^2 f^{-4}(\rho - f)[z_R^2 + (a - f)^2] - 2\Delta f^{-2}[z_R^2 + (a - f)(2\rho + a - 3f)] - (2f - a - \rho)\} - (i z_R)}{(2\Delta f^{-2}(\rho - f) - 1)^2 z_R^2 + (2\Delta f^{-2}(f - a)(f - \rho) + 2f - a - \rho)^2}. \tag{6}$$

Equating the imaginary parts,

$$\frac{\pi w^2}{\lambda_o} = (2\Delta f^{-2}(\rho - f) - 1)^2 z_R + (2\Delta f^{-2}(f - a)(f - \rho) + 2f - a - \rho)^2 \frac{1}{z_R}. \tag{7}$$

Thus, the equation for the beam spot size $w^2$ is

$$w^2 = (2\Delta f^{-2}(\rho - f) - 1)^2 w_o^2 + (2\Delta f^{-2}(f - a)(f - \rho) + 2f - a - \rho)^2 \frac{\lambda_o^2}{\pi^2 w_o^2}, \tag{8}$$

where $a$ is the distance between the laser source and the objective lens; $\rho$ is the distance from the image sensor to the objective lens; $R$ is the wavefront radius of the laser beam at the focus, which is supposed to be infinite ($R \to \infty$); and $f$ is the focal length of the objective lens. If the specimen is located at the focus, the beam spot radius $w$ is

$$w = \sqrt{w_o^2 + (a + \rho - 2f)^2 \frac{\lambda_o^2}{\pi^2 w_o^2}}. \quad (9)$$

Equation (9) indicates that $w$ at the image sensor is a function of the distances, $f$, and the characteristic parameters of the laser. To optimize the images collected by the image sensor, the distances are the main factors that should be varied, as long as high-resolution beam spot images can be obtained by the image sensor. Furthermore, it can be seen that as $f$ becomes longer, $w$ smaller becomes if the specimen is located at the focus. Consequently, it is easier to determine the coincidence of the beam spots before and after the blocking plate is removed during calibration while moving the specimen away from the focal position, which will also result in more accurate measurements. The Fig.2 indicates the image of dental sample under the high-precision laser ablation when the decayed enamel region is completely removed with support of the focus inspection system. The dental ultrashort pulsed laser used to obtain this result holds the energy of 16 $\mu$J with the repetition rate of 500 kHz, focal spot radius of 20 $\mu$m, and wavelength of 1064 nm without water cooling. According to the figure, the ablation hole has the circle shape of focal beam spot with acceptable depth and considerable uniformity. While the decayed region is smoothly ablated, the surrounding region is perfectly maintained. This fact contributes to reducing the pain, side effect, and sequela after laser operation, and maintaining the beauty and function of the dental structures.

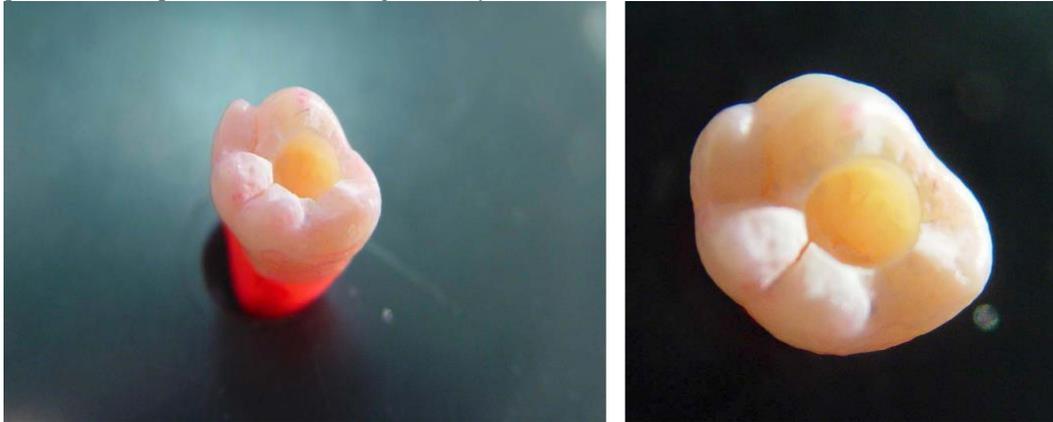

Fig.2: Result of laser ablation for caries removal.

### 4. Conclusion

We proposed a novel technique for detecting the focal position of a target sample during dental laser operation for decay removal by examining the beam spot configuration and treating the laser beam as a stream of photons undergoing quantum correlation. Defocusing distance of the sample surface from the focus can be computed via spatial correlation of photons hitting PSDs. Potentially, the precision of this technique can reach the Heisenberg limit, providing a new approach for designing high-accuracy automatic focus detection systems in the future. The proposed method is novel because it combines the principles of quantum optics and geometrical optics for focusing condition manipulation in medical laser technology.